\documentstyle[manuscript,aps,prbbib,graphicx]{revtex}

\tighten
\newcommand{\be}{\begin{equation}}
\newcommand{\ee}{\end{equation}}
\newcommand{\bea}{\begin{eqnarray}}
\newcommand{\eea}{\end{eqnarray}}
\newcommand{\bd}{\begin{displaymath}}
\newcommand{\ed}{\end{displaymath}}
\newcommand{\bi}{\begin{itemize}}
\newcommand{\ei}{\end{itemize}}
\newcommand{\bc}{\begin{center}}
\newcommand{\ec}{\end{center}}
\newcommand{\bfl}{\begin{flushleft}}
\newcommand{\efl}{\end{flushleft}}
\newcommand{\bfr}{\begin{flushright}}
\newcommand{\efr}{\end{flushright}}
\newcommand{\f}{\frac}

\def\bk{{\bf k}} \def\bq{{\bf q}} \def\bp{{\bf p}}
  
\def\ra{\rightarrow}
\def\6{\partial} \def\a{\alpha} \def\b{\beta}
 \def\d{\delta} \def\ve{\varepsilon} 
\def\z{\zeta}  \def\th{\theta}
  \def\l{\lambda}

\def\o{\omega} \def\G{\Gamma} \def\D{\Delta}
  
  \def\O{\Omega}

\def\={\!\!\!&=&\!\!\!}
\def\+{\!\!\!&&\!\!\!+~}
\def\-{\!\!\!&&\!\!\!-~}

\begin{document}

\title{Electron-fluctuation interaction in a non-Fermi superconductor}
\author{M. Crisan}
\address{Department of Theoretical Physics, University of Cluj,
3400 Cluj, Romania}
\author{C. P. Moca}
\address{Department of Physics, University of Oradea,
3700 Oradea, Romania}
\author{I. Tifrea}
\address{Department of Theoretical Physics, University of Cluj,
3400 Cluj, Romania}
\maketitle

\begin{abstract}
We studied the influence of the amplitude fluctuations of a non-Fermi
superconductor on the energy spectrum of the 2D Anderson non-Fermi system.
The classical fluctuations give a temperature dependence
in the pseudogap induced in the fermionic excitations.
\end{abstract}
\newpage
\section*{Introduction}
The microscopic description of the superconducting state in cuprate materials
is a very difficult problem because at the present time is generally accepted
that in the normal state the elementary excitations are not described by the
Fermi liquid theory. However, using the BCS-like pairing model the Gorkov
equations have been applied to describe the superconducting state in the
hypothesis that the normal state is a non-Fermi liquid described by the
Anderson model \cite{1}. The superconducting state properties have been
discussed by different authors \cite{2,3,4,5,6,7,8} and even if these
descriptions are phenomenological, they can be a valid starting point for a
microscopic model. Recent experimental data (ARPES) showed that these
materials present even more remarkable deviations from the Fermi liquid
behavior due to the occurrence of the pseudogap at the Fermi surface.

The occurrence of the pseudogap has been explained using different concepts
as: the spin fluctuations \cite{9}, preformed pairs \cite{10}, SO(5) symmetry
\cite{11}, spin-charge separation \cite{12}, the fluctuations of the order
parameter induced pseudogap \cite{13}.

In this paper we start with a non-Fermi liquid description of the
superconducting state (See Ref. \onlinecite{1,2,3,4,5,6,7,8}) and consider
the interaction between the order parameter fluctuations and the electrons
(Section 2 and Section 3).
This problem has been studied by Abrahams et al. \cite{14}, Marcelja \cite{15}
and Schmid \cite{16} for BCS superconductors and the theory explained the
tunneling experiments on films, by the modification of the density of states
by a pseudogap which appears at a temperature higher than the BCS critical
temperature. Using such an approximation we will calculate (Section 4) the
pseudogap due to the electron-fluctuation interaction and in the simple
mode-mode approximation the temperature dependence of it will be obtained.

Finally (Section 5) we compare our results with the other theoretical models
for the cuprate superconductors.

\section*{The model}
The non-Fermi behavior of the normal state for the cuprate superconductors
proposed by Anderson \cite{1} was developed by different authors
\cite{2,3,4,5,6,7,8} in order to describe the superconducting state in the
framework of the BCS theory. In the normal state the electrons are described
by the Green's function
\be
G_0(\bk,i\o_n)=\f{\o_c^{-\a}}{(i\o_n-\ve_\bk)^{1-\a}}
\label{e1}
\ee
where $\o_c$ is a cutoff energy and $0<\a<1$.

In the following we consider that the superconducting state appears due an
attractive interaction and is described by the BCS like order parameter
$\D_\bk$ which can be calculated from the Gorkov equations. The fluctuations
of this parameter can interact with the electrons and the fermionic spectrum
of the elementary excitations changes. Such an effect has been studied in
the BCS superconductors by different authors \cite{13,14,15} and it was showed
that this interaction gives a contribution to the density of states for
$T>T_c$ which explained the behavior of the tunneling measurements.

For a superconductor described by the Gorkov like equations with the normal
state described by Eq. (\ref{e1}) the propagator of the fluctuations has the
expression:
\be
D(\bq, i\o_n)=\f{1}{V^{-1}+\Pi(\bq, i\o_n)}
\label{e2}
\ee
where $V$ is the attractive interaction between the electrons and
$\Pi(\bq, i\o_n)$ is the polarization operator defined as
\be
\Pi(\bq, i\o_n)=T\sum_{\o_l}\int\f{d\bp}{(2\pi)^2}G(\bp, i\o_l)
G(\bq-\bp, i\o_n-i\o_l)
\label{e3}
\ee
where $G(\bp, i\o_l)$ is the Green's function related to electrons, which in
terms of a Dyson equation has the following form
\be
G^{-1}(\bp, i\o_l)=G_0^{-1}(\bp, i\o_l)-\Sigma(\bp, i\o_l)
\label{e5}
\ee
where the self energy is given by
\be
\Sigma(\bp, i\o_l)=-T\sum_{\o_n}\int\f{d\bq}{(2\pi)^2}D(\bq, i\o_n)
G(\bq-\bp, i\o_n-i\o_l)
\label{e4}
\ee
Eqs. (\ref{e2}-\ref{e4}) have to be solved self consistent, but this cannot be
done analytically. However, in the mode-coupling approximation it can be done
and we can calculate the new energy of the electronic excitations.

\section*{Mode-coupling approximation}
In this approximation we consider first that $G(\bk, i\o_n)\approx
G_0(\bk, i\o_n)$ and from Eq. (\ref{e3}) we define the polarization
\be
\Pi_0(\bq, i\o_m)=\int\f{d\bk}{(2\pi)^2} S(\bk, \bq, i\o_m)
\label{e6}
\ee
where
\be
S(\bk, \bq, i\o_m)=(-1)^{1-\a}T\sum_{\o_n}\f{\o_c^{-\a}}
{(i\o_n-\ve_\bk)^{1-\a}(i\o_n-i\o_m+\ve_{\bq-\bk})^{1-\a}}
\label{e7}
\ee
We performed the analytical calculation of $\Pi_0(\bq, i\o_m)$ given by
Eq. (\ref{e6}) (See Appendix) and from Eq. (\ref{e2}) the propagator for the
order parameter fluctuations has been obtained as
\bea
D_0^{-1}(\bq, i\o_n)&=&N(0) A(\a)\left\{C(\a)
\left[\left(\f{T}{\o_c}\right)^{2\a}-\left(\f{T_c}{\o_c}\right)^{2\a}\right]
+\f{i\o_n(1-\a)}{T} M\left(\a,\f{T}{\o_c},\f{\o_D}{\o_c}\right)\right.\nonumber\\
&+&\left.\left(\f{v_Fq}{2T}\right)^2(1-\a)^2 N\left(\a,\f{T}{\o_c},
\f{\o_D}{\o_c}\right)\right\}
\label{e8}
\eea
where the critical temperature $T_c$ has been obtained \cite{7,8} as
\be
T_c^{2\a}=\f{1}{C(\a)}\left[D(\a)\o_D^{2\a}-\f{\o_c^{2\a}}{A(\a)N(0)V}\right]
\label{e9}
\ee
and the constants from Eqs. (\ref{e8}) and (\ref{e9}) are
\bd
A(\a)=\f{2^{2\a}}{\pi}\sin{\pi (1-\a)}
\ed
\bd
C(\a)=\G^2(\a)\left[1-2^{1-2\a}\right]\zeta(2\a)
\ed
\bd
D(\a)=\f{\G(1-2\a)\G(\a)}{2\a\G(1-\a)}
\ed
\bea
M\left(\a,\f{T}{\o_c},\f{\o_D}{\o_c}\right)&=&\f{\G(\a-1)\G(\a-1/2)}{2\sqrt{\pi}}
\left[1-2^{2-2\a}\right]\zeta(2\a-1)\left(\f{T}{\o_c}\right)^{2\a}\nonumber\\
&-&\f{B(3-2\a,\a-2)}{4(2\a-2)}\left(\f{\o_D}{\o_c}\right)^{2\a-2}
\left(\f{T}{\o_c}\right)^2\\
N\left(\a,\f{T}{\o_c},\f{\o_D}{\o_c}\right)&=&
\left[\f{2\G(\a-2)\G(\a-1/2)}{\sqrt{\pi}}+\G^2(\a)\right]\f{1-2^{3-2\a}}{4}
\zeta(2-2\a)\left(\f{T}{\o_c}\right)^{2\a}\nonumber\\
&-&\f{B(3-2\a,\a-2)}{4(2\a-2)}\left(\f{\o_D}{\o_c}\right)^{2\a-2}
\left(\f{T}{\o_c}\right)^2\nonumber
\eea
$B(x,y)=\G(x)\G(y)/\G(x+y)$ and $\G(x)$ are the Euler' functions and
$\zeta(x)$ is the Riemann function.

Using a similar form with the one introduced by Schmid the fluctuation
propagator will be written as
\be
D_0^{-1}(\bq, i\o_n)=N(0)\left[b(\a)\tau(\a)+ia(\a)\o_n+\xi^2(\a,T)q^2\right]
\label{e10}
\ee
where
\be
\tau(\a)=\left(\f{T}{T_c}\right)^{2\a}-1
\label{e11}
\ee
\be
a(\a)=\f{M(\a,T/\o_c,\o_D/\o_c)}{T}(1-\a)A(\a)
\label{e13}
\ee
\be
b(\a)=A(\a)C(\a)\left(\f{T_c}{\o_c}\right)^{2\a}
\label{e14}
\ee
and
\be
\xi(\a)=\f{v^2_F(1-\a)^2}{4T^2} N\left(\a,\f{T}{\o_D},\f{\o_D}{\o_c}\right)
A(\a)
\label{e12}
\ee

In the approximation $\Sigma\ll \pi T$ the Green function given by
Eq. (\ref{e5}) will be approximated as $G=G_0+G_0\Sigma G_0$ and $\Pi$ will
be modified by $\d\Pi$ also linear in $\Sigma$. Following Ref. \onlinecite{11}
we calculated $\d\Pi$ in the "box approximation" as
\be
\d\Pi=2T^2\sum_n\int\f{d\bp}{(2\pi)^2}G_0^2(\bp, i\o_n)G_0^2(-\bp,-i\o_n)
\int\f{d\bq}{(2\pi)^2}D(\bq,\o_n=0)
\label{e15}
\ee
where
\be
D^{-1}(\bq, i\o_n)=\f{1}{V}+\Pi(\bq,i\o_n)+\d\Pi(\bq,i\o_n)
\label{e16}
\ee
In order to calculate $\d\Pi$ we introduce
\be
B_0=\f{1}{N(0)}T\sum_n\int\f{d\bp}{(2\pi)^2}G_0^2(\bp,i\o_n)G_0^2(-\bp,-i\o_n)
\label{e17}
\ee
where $N(0)=m/2\pi$. If we use for the electronic Green function Eq. (\ref{e1})
we obtained
\be
B_0(T)=\f{B(1/2,3/2-2\a)}{\pi}\f{\left[2^{3-4\a}-1\right]\zeta(3-4\a)}
{2^{3-4\a}}\f{\o_c^{-4\a}}{(\pi T)^{2-4\a}}
\label{e19}
\ee
If we introduce $\tilde{\tau}(\a)=\tau(\a)+\d\Pi/N(0)$
the fluctuation propagator given by Eq. (\ref{e16}) will be
\be
D^{-1}(\bq, i\o_n)=b(\a)\tilde{\tau}(\a)+ia(\a)\o_n+\xi^2(\a,T)q^2
\label{e21}
\ee
where
\be
\tilde{\tau}(\a)-\tau(\a)=\f{2B_0(T)}{N(0)} T\int\f{\bq}{(2\pi)^2}
\f{1}{\tilde{\tau}(\a)+\xi^2(\a,T)q^2}
\label{e22}
\ee
If we perform this integral taking the upper limit $q_M=1/\xi(\a,T)$ from
Eq. (\ref{e22}) we get
\be
\tilde{\tau}(\a)-\tau(\a)=\f{B_0(T) T}{2\pi N(0)\xi^2(\a,T)}
\ln{\f{1+\tilde{\tau}(\a)}{\xi^2(\a,T)}}
\label{e23}
\ee
For realistic parameters ($T_c=100K$, $\o_c=200K$) the difference
$\tilde{\tau}(\a)-\tau(\a)$ becomes important only near a critical value of
$\a$ defined by $\xi(\a_c)=0$. In the BCS limit ($\a=0$) this parameter is
small and this behavior can be associated with the occurrence of the
preformed pairs in the domain $T_c<T<T^*$, controlled by $\a$. This behavior
is in fact due to the occurrence of a pseudogap in the electronic
excitations.

\section*{Electronic self-energy}
The self-energy due to the interaction between electrons and fluctuations
is given by Eq. (\ref{e4}) where $D(\bq,i\o_n)$ is given by Eq. (\ref{e21}).
First we calculate the summation over the Matsubara frequencies $\o_n$
\be
S=T\sum_n D(i\o_n)G(i\o_n-i\o_l)=T\sum_n\f{(-1)^\a\o_c^{-\a}}
{N(0)(b\tilde{\tau}+ia\o_n+\xi^2q^2)(i\o_l+\ve_\bk-i\o_n)^{1-\a}}
\label{e24}
\ee
transforming this sum in a contour integral which has a pole at
$\O(\bq)=-(b\tilde{\tau}+\xi^2q^2)/a$ and a cut line from $\ve_\bk+i\o_l$
to $\infty$ in the upper semiplane. From Eq. (\ref{e13}) we can see that
$a(\a)=-|a(\a)|$ and in fact $\O(\bq)=(b\tilde{\tau}+\xi^2q^2)/|a|$.
Performing this integral we obtain
\bea
S&=&\f{\o_c^{-\a}}{N(0)}\f{n(\O(\bq))}{(-i\o_l-\ve_\bk-\O(\bq))^{1-\a}}
\nonumber\\
&-&\f{\o_c^{-\a}}{N(0)}\f{\sin{[\pi (1-\a)]}}{\pi}\int_{\ve_\bk}^\infty dt
\f{f(t)}{(b\tilde{\tau}-|a|(t+i\o_l)+\xi^2q^2)(t-\ve_\bk)^{1-\a}}
\label{e25}
\eea
where $n(x)$ is the Bose-Einstein function and $f(x)$ is the Fermi-Dirac
function and $\ve_\bk=k^2/2m-E_F$. The integral from the second contribution
in Eq. (\ref{e25}) will be performed using the expansion
$f(t)=\sum_{m=0}(-1)^m\exp{[-\b (m+1)t]}$ and the last term becomes
\bea
I_1&=&\sum_{m=0}^\infty\f{(-1)^m}{|a|}\f{(\ve_\bk+\O(\bq))^{\a/2+1}}
{[\b(m+1)]^\a/2}\exp{\left[\f{\b(m+1)(\O(\bq)-\ve_\bk)}{2}\right]}\nonumber\\
&\times&\G(\a)W_{-\a/2,\a/2-1/2}\left[\b(m+1)(\O(\bq)+\ve_\bk)\right]
\label{e26}
\eea
where the Whittaker function $W_{\l,\mu}(z)$ will be approximated as
$W_{\l,\mu}\cong e^{-z/2}z^\l$. This results give for Eq. (\ref{e25}) the
expression
\bea
S&=&\f{\o_c^{-\a}}{N(0)}\f{n(\O(\bq))}{(-i\o_l-\ve_\bk-\O(\bq))^{1-\a}}
\nonumber\\
&+&\f{\o_c^{-\a}}{N(0)}\f{\sin{[\pi(1-\a)]}}{\pi}
\sum_{m=0}^\infty\f{(-1)^m}{|a|}\f{\ve_\bk+i\o_l+\O(\bq)}{[\b(m+1)]^\a}
\G(\a)\exp{[\b(m+1)\ve_\bk]}
\label{e27}
\eea
In the limit $k\cong k_F$ the second term denoted by $S_2$ becomes
\be
S_2=\f{\o_c^{-\a}}{N(0)}\f{\sin{[(1-\a)\pi]}}{\pi}
\f{i\o_l|a|+b\tilde{\tau}+\xi^2q^2}{|a|^2}\G(\a)[1-2^{1-\a}]\zeta(\a)
\label{e28}
\ee
and if $T\ra T_c$, $\o_l\ra 0$ and $q\ra 0$ this term can be neglected. This
approximation is in fact equivalent with the physical picture proposed by Vilk
and Tremblay \cite{13} in which the occurrence of the pseudogap is given by
the interaction between the electrons and the classical fluctuations. Indeed,
in this regime the first term of Eq. (\ref{e27}) can be written as
\be
S\cong \f{1}{N(0)}n(\O(\bq))G(\bk,-i\o_l+\O(\bq))
\label{e30}
\ee
and the electronic self-energy becomes
\be
\Sigma(\bp,\o+i0)\cong -\D_{pg}^2G(\bk,-i\o_l)
\label{e31}
\ee
where we considered $\ve_\bk\gg \O(\bq)$ and
\be
\D^2_{pg}=\f{1}{N(0)|a|}\int\f{d\bq}{(2\pi)^2} n(\O(\bq))
\label{e32}
\ee
will be approximated as
\be
\D_{pg}^2(T)\cong \f{T}{4\pi|a|N(0)}\int_0^{q_M}\f{q dq}{(\tilde{\tau}+\xi^2q^2)/|a|}
\label{e33}
\ee
where $q_M$ is the wave number cutoff. From Eq. (\ref{e33}) we calculate the
temperature dependence of $\D_{pg}(T)$ as
\be
\D_{pg}^2(T)=\f{T}{4\pi N(0)\xi^2}\ln\left(1+\f{\xi^2}{\tilde{\tau}}q_M^2\right)
\label{e34}
\ee
\section*{Discussions}

We showed that a temperature dependent pseudogap appears in a non-Fermi
superconductor due to the interaction between electrons and the
fluctuations of the order parameter amplitude.

The mode-mode coupling, valid in the weak coupling approximation, can give
relevant results, even for the intermediate coupling studied by the Levin
group \cite{17} using the resonant scattering model. The method, recently
applied by Norman et al \cite{18}, Randeria \cite{19} can be applied for
the spin-fluctuation model proposed by Chubukov \cite{9} in order to study
the temperature dependence of the pseudogap. In Ref. \onlinecite{18} and
\onlinecite{19} the filling in of the pseudogaps due to the incresment of
the temperature is given by the broadening in the self-energy and is
proportional to $T-T_c$. A similar broadening effect, proportional to
$\tilde{\tau}(\a)$ was obtained in our model and this can be seen very easy
from Eq. (\ref{e28}) if in the electronic Green function we take the limit
$q=0$.

Recently such a model, for a Fermi liquid superconductor has been studied
by Kristoffel and Ord \cite{20} and their temperature dependence is
different from our result. However, we mention that according to their model
these authors have to obtain a result similar to the result given in \cite{13}.
The difference is given by the method of performing the integral over $\bq$
which is not correct in \cite{19}.

Recently, Preotsi et al \cite{21} generalized the method given in \cite{13}
taking into consideration the anysotropy in the dynamic susceptibility due
to the interplane pairing.

\section{Acknowledgments}
We thank Mohit Randeria and Andre-Marie Tremblay for useful discussions
about recent developments in pseudogap models. The work was supported by
MEI under the grant nr. 184/1998.

\section*{Appendix}
The polarization $\Pi(\bq,i\o_m)$ for a 2D non-Fermi liquid is defined as
\bea
\Pi_0(\bq, i\o_m)&=&T\sum_n\int\f{d\bk}{(2\pi)^2}G_0(\bk, i\o_n)
G_0(\bq-\bk,i\o_m-i\o_n)\nonumber\\
&=&T\sum_n\int\f{d\bk}{(2\pi)^2}\f{\o_c^{-2\a}}{(i\o_n-\ve_\bk)^{1-\a}
(i\o_m-i\o_n-\ve_{\bq-\bk})^{1-\a}}
\label{a1}
\eea
which can be written as
\be
\Pi_0(\bq,i\o_m)=\int\f{d\bk}{(2\pi)^2}S(\bk,\bq,i\o_m)
\label{a2}
\ee
where
\be
S(\bk,\bq,i\o_m)=(-1)^{1-\a}T\sum_n\f{\o_c^{-2\a}}
{(i\o_n-\ve_\bk)^{1-\a}(i\o_n+\ve_{\bq-\bk}-i\o_m)^{1-\a}}
\label{a3}
\ee
In order to perform the summation in Eq. (\ref{a3}) we transform he summation
in a contour integral
\be
S(\bk,\bq,z)=-\oint_C\f{dz}{2\pi i}n(z)F(z)
\label{a4}
\ee
where $n(z)$ is the Fermi function and $F(z)$ is given by
\be
F(z)=\f{(-1)^{1-\a}\o_c^{-2\a}}{(z-\ve_\bk)^{1-\a}
(z+\ve_{\bq-\bk}-i\o_m)^{1-\a}}
\label{a5}
\ee
and the contour C is taken as $(-\infty,i\o_n+\ve_{\bq-\bk})\bigcup
(\ve_\bk,\infty)$. The integral in Eq. (\ref{a4}) has been evaluated as
\bea
\oint_C\f{dz}{2\pi i}n(z)F(z)&=&\f{1}{2\pi i}\left\{
\int_{-\infty}^{-\ve_{\bq-\bk}}dx n(x+i\o_m)\f{2i\o_c^{-2\a}\sin{[\pi(1-\a)]}}
{(x-\ve_\bk)^{1-\a}(x+i\o_n+\ve_{\bq-\bk})^{1-\a}}\right.\nonumber\\
&-&\left.\int_{\ve_\bk}^\infty dx n(x) \f{2i \o_c^{-2\a}\sin{[\pi(1-\a)]}}
{(x-\ve_\bk)^{1-\a}(x+i\o_m+\ve_{\bq-\bk})^{1-\a}}\right\}
\label{a6}
\eea
In order to perform the integral over $x$ we express the dominators from
(\ref{a6}) as
\bea
(x-\ve_\bk)^{\a-1}(x+\ve_{\bq-\bk}+i\o_n)^{\a-1}&=&(x-\ve_\bk)^{\a-1}
(x+\ve_\bk)^{\a-1}\nonumber\\
&-&(\a-1)(v_Fq\cos{\theta}-i\o_n)(x-\ve_\bk)^{\a-1}(x+\ve_\bk)^{\a-2}\nonumber\\
&+&\f{(\a-1)^2}{2}(v_Fq\cos{\theta})^2(x-\ve_\bk)^{\a-1}(x+\ve_\bk)^{\a-3}
\label{a7}
\eea
and take for the Fermi function the expansion
\be
n(x)=\sum_{m=0}^\infty(-1)^m\exp{[-\b (m+1)x]}
\label{a8}
\ee
Using now the integrals
\be
\int_u^\infty dx (x+\b)^{-\nu}(x-u)^{\mu-1}=(u+\b)^{\mu-\nu}B(\nu-\mu,\mu)
\label{a9}
\ee
\bea
\int_u^\infty dx (x+\b)^{2\nu-1}(x-u)^{2\rho-1}\exp{[-\mu x]}&=&
\f{(u+\b)^{\nu+\rho+1}}{\mu^{\nu+\rho}}\exp{\left[\f{(\b-u)\mu}{2}\right]}\nonumber\\
&\times&\G(2\rho)W_{\nu-\rho,\nu+\rho-1/2}(u\mu+\b\mu)
\label{a10}
\eea
we calculated $S(\bk,\bq,i\o_m)$ as
\bea
S(\bk,\bq,\o_m)&=&\f{\o_c^{-2\a}\sin{[\pi (1-\a)]}}{\pi}\nonumber\\
&\times&\left\{-(2\ve_\bk)^{2\a-1}B(1-2\a,\a)\right.\nonumber\\
&+&\f{2\G(\a)}{\sqrt{\pi}}\sum_{m=0}^\infty (-1)^m
\f{(2\ve_\bk)^{\a-1/2}}{[\b(m+1)]^{\a-1/2}}K_{\a-1/2}[\ve_\bk\b(m+1)]\nonumber\\
&+&(\a-1)(v_Fq\cos{\th}-\o_m)\left[-(2\ve_\bk)^{2\a-2}B(2-2\a,\a-1)\right.\nonumber\\
&+&\left.\f{\G(\a-1)}{\sqrt{\pi}}
\sum_{m=0}^\infty (-1)^m\f{(2\ve_\bk)^{\a-1/2}}{[\b(m+1)]^{\a-3/2}}
K_{\a-3/2}[\ve_\bk\b(m+1)]\right]\nonumber\\
&+&\f{(\a-1)^2(v_Fq\cos{\th})^2}{2}\left[-(2\ve_\bk)^{2\a-3}B(3-2\a,a-2)\right.\nonumber\\
&+&\f{\G(\a-2)}{\sqrt{\pi}}\sum_{m=0}^\infty (-1)^m\f{(2\ve_\bk)^{\a-1/2}}{[\b(m+1)]^{\a-5/2}}
K_{\a-3/2}[\ve_\bk\b(m+1)]\nonumber\\
&+&\left.\left.\f{2\G(\a-1)}{\sqrt{\pi}}\sum_{m=0}^\infty (-1)^m\f{(2\ve_\bk)^{\a-3/2}}{[\b(m+1)]^{\a-3/2}}
K_{\a-3/2}[\ve_\bk\b(m+1)]\right]\right\}
\label{a11}
\eea
Eq. (\ref{a2}) will be written as
\be
\Pi(\bq,i\o_m)=2N(0)\int_0^{2\pi}\f{d\theta}{2\pi}\int_0^{\o_D}d\ve
S(\ve,\bq,i\o_m)
\label{a12}
\ee
and using the relation
\bd
\int_0^\infty x^\mu K_\nu(ax)dx=2^{\mu-1}a^{-\mu-1}
\G\left(\f{1+\mu+\nu}{2}\right)\G\left(\f{1+\mu-\nu}{2}\right)
\ed
we obtained
\bea
\Pi_0(\bq, \o)&=&\f{2N_0\sin{[\pi(1-\a)]}}{\pi}\nonumber\\
&\times&\left\{-\f{2^{2\a-1}B(1-2\a,\a)}{2\a}\left(\f{\o_D}{\o_c}\right)^{2\a}+
\G^2(\a)\f{1-2^{1-2\a}}{2^{1-2\a}}\z(2\a)\left(\f{T}{\o_c}\right)^{2\a}\right.\nonumber\\
&+&\f{i\o(1-\a)}{\o_c}\left[-\f{2^{2\a-2}B(2-2\a,\a-1)}{2\a-1}
\left(\f{\o_D}{\o_c}\right)^{2\a-1}\right.\nonumber\\
&+&\left.\f{\G(\a-1)\G(\a-1/2)}{\sqrt{\pi}}\f{1-2^{2-2\a}}{2^{2-2\a}}\z(2\a-1)
\left(\f{T}{\o_c}\right)^{2\a-1}\right]\nonumber\\
&+&\f{(v_Fq)^2(1-\a)^2}{4\o_c^2}\left[-\f{2^{2\a-3}B(3-2\a,\a-2)}{2\a-2}
\left(\f{\o_D}{\o_c}\right)^{2\a-2}\right.\nonumber\\
&+&\left.\left.\left(\f{2\G(\a-2)\G(\a-1/2)}{\sqrt{\pi}}+\G^2(\a-1)\right)\f{1-2^{3-2\a}}{2^{3-2\a}}
\z(2-2\a)\left(\f{T}{\o_c}\right)^{2\a-2}\right]\right\}
\label{a13}
\eea
Using now the Thouless criterion
\be
1+V Re \Pi(\bq=0,i\o_m=0)=0
\label{a14}
\ee
we calculate the critical temperature
\bea
\left(\f{T_c}{\o_c}\right)^{2\a}&=&\f{B(1-2\a,\a)}{2\a\G^2(\a)(1-2^{1-2\a})
\zeta(2\a)}\left(\f{\o_D}{\o_c}\right)^{2\a}\nonumber\\
&-&\f{\pi}{N(0)V\G^2(\a)2^{2\a}(1-2^{1-2\a})\zeta(2\a)\sin{[\pi(1-\a)]}}
\label{a15}
\eea
which is identical to Eq. (\ref{e8}) if we introduce $\l=N(0)V$.

\end{document}